\documentclass[reprint,showpacs,twocolumn,superscriptaddress,aps,prb]{revtex4-2}
\usepackage[T1]{fontenc}        
\usepackage[utf8]{inputenc}
\usepackage[english]{babel}

\usepackage[protrusion,     expansion]{microtype} 
            
\usepackage{mathtools}
\usepackage{amsfonts, amsmath, amsthm, amssymb, mathrsfs}
\usepackage{graphicx}   
\usepackage[usenames,dvipsnames]{xcolor}    
\usepackage{braket}
\usepackage{tikz}
\usepackage{url}
\usepackage{ulem}

\usepackage{appendix}
\usepackage{fancyhdr}

\usepackage{braket}
\usepackage{xspace}
\usepackage{mathdots}
\usepackage{stackrel}
\usepackage{xfrac}

\usepackage{bbm}      

\definecolor{mycolor3}{HTML}{BF1363}

\usepackage[colorlinks,
            linkcolor=black,
            citecolor=black,
            urlcolor=mycolor3,
            bookmarks,
            bookmarksnumbered
            ]{hyperref}    

\usepackage[capitalize]{cleveref}
\crefname{figure}{Fig.}{Figs.}
\crefname{equation}{Eq.}{Eqs.}
\crefname{section}{Sec.}{Secs.}

\allowdisplaybreaks

\theoremstyle{definition}

\theoremstyle{remark}

\newcommand{\ii}{\mathrm{i}}

\newcommand{\dd}{\mathrm{d}}

\newcommand{\abs}[1]{\left\vert#1\right\vert}

\DeclareMathOperator{\Tr}{Tr}   

\begin{document}

\title{Entanglement Hamiltonians and the quasiparticle picture}
\author{Federico Rottoli}
\affiliation{SISSA and INFN Sezione di Trieste, via Bonomea 265, 34136 Trieste, Italy.}
\affiliation{Dipartimento di Fisica dell’Universit\`a di Pisa and INFN Sezione di Pisa, Largo B. Pontecorvo 3, 56127 Pisa, Italy.}
\author{Colin Rylands}
\affiliation{SISSA and INFN Sezione di Trieste, via Bonomea 265, 34136  Trieste, Italy.}
\author{Pasquale Calabrese}
\affiliation{SISSA and INFN Sezione di Trieste, via Bonomea 265,   34136 Trieste, Italy.} 
\affiliation{International Centre for Theoretical Physics (ICTP), Strada Costiera 11, 34151 Trieste, Italy.}

\begin{abstract}
The entanglement Hamiltonian (EH) provides the most comprehensive characterization of bipartite entanglement in many-body quantum systems. Ground states of local Hamiltonians inherit this locality, resulting in EHs that are dominated by local, few-body terms. 
Unfortunately, in non-equilibrium situations, analytic results are rare and largely confined to continuous field theories, which fail to accurately describe microscopic models. 
To address this gap, we present an analytic result for the EH following a quantum quench in non-interacting fermionic models, valid in the ballistic scaling regime.
The derivation adapts the celebrated quasiparticle picture to operators, providing detailed insights into its physical properties.
The resulting analytic formula serves as a foundation for engineering EHs in quantum optics experiments.
\end{abstract}
\maketitle

\paragraph*{Introduction.}
The study of bipartite entanglement in quantum many-body systems has revolutionized our understanding of quantum physics, providing profound insights into the behavior of complex systems across various fields, from quantum information~\cite{nc-10,eisert-2010} and condensed matter~\cite{intro1,intro2,intro3} to high-energy physics~\cite{nrt-09,hol}.
Entanglement Hamiltonians (EHs), which describe the reduced density matrix (RDM) of a subsystem as an effective Hamiltonian, have emerged as a powerful tool in this domain~\cite{w-18,defv-22}. They represent the most complete characterization of bipartite entanglement at an operatorial level.
For a given state $\rho$, the RDM of a region $A$, $\rho_A$, is obtained by tracing $\rho$ over the complement $\bar A$ of $A$, that is
\begin{equation}
   \rho_A=\Tr_{\bar A} \rho=\frac{e^{-K_A}}{Z_A}\,, \quad Z_A = \Tr e^{-K_A}\,,
\end{equation}
where the operator $K_A$ is the EH.
The RDM is usually a very complicated and full matrix. 
It is therefore quite surprising that for the ground states of relativistic quantum field theories, the celebrated Bisognano-Wichmann theorem~\cite{bw-75,bw-76} ensures that the EH has a very local structure. 
Such locality is reflected in lattice many-body systems in which the EH is dominated by local few-body terms~\cite{it-87,ep-17,ep-18,ETP19,EH-1,EH-2,jt-21,gmcd-18,zcdr-20,fsc-22,rfc-24,trs-18,etp-22}.
Aside from being interesting from a theoretical perspective, the locality of the EH is also a key feature that underpins its experimental implementation~\cite{dvz-17,kbevz-21,var-2,exp1}. 

It is thus very natural to explore and search for other physical situations where the EH is local. To date, local EHs have primarily been derived in the realm of conformal field theory (CFT), where conformal transformations enable the mapping of ground state results to more complex scenarios~\cite{hl-82,chm-11,kw-13,ct-16}, such as quantum quenches~\cite{ct-16,wrl-18}.
Notwithstanding some studies on EHs after quantum quenches in lattice models~\cite{gat-19,zhhw-20}, it is still unclear to what extent the locality of the EH is preserved in the absence of conformal symmetry. 
The goal of this Letter is to initiate a systematic study of the EHs in non-equilibrium situations. 
We focus here on free fermionic models and show that in general the EH after a global quantum quench can be reconstructed from the quasiparticle picture (QPP)~\cite{cc-05,ac-16,ac-17c,c-18}.
Originally developed to describe entanglement entropy growth after a quantum quench in integrable systems, this picture posits that entanglement is carried by pairs of quasiparticles emitted from the initial state.  As these quasiparticles propagate through the system, they spread entanglement in a manner that can be quantitatively tracked and predicted.  
It has since been shown to be applicable to the calculation of other quantities such as some correlation functions~\cite{cef-12}, negativity~\cite{ctc-14,ac-18b}, full counting statistics~\cite{gec-18,hr-24,bcckr-22}, symmetry resolved entanglement~\cite{bcckr-22,pbc-20,pbc-21b}, operator entanglement~\cite{dubail,rvm-22}, and the entanglement asymmetry~\cite{amc-22,makc-23,bkccr-23}.
However, despite the fact that it contains information about most of these quantities, the application of the QPP to describe the EH has remained extremely elusive.
In particular, while the R\'enyi entropies indicate that the spectrum of the RDM admits a QPP it is far from trivial that the matrix itself does.
In this letter, we fill this void and derive a very compact form for the EH after a generic integrable quench in a free-fermionic model.
We proceed by first presenting the result, followed by its derivation and then present some analytic and numerical checks of its veracity.

\paragraph*{Main result for the post-quench Entanglement Hamiltonian.}

\begin{figure*}[t]
\centering
\includegraphics[width=0.38\textwidth]{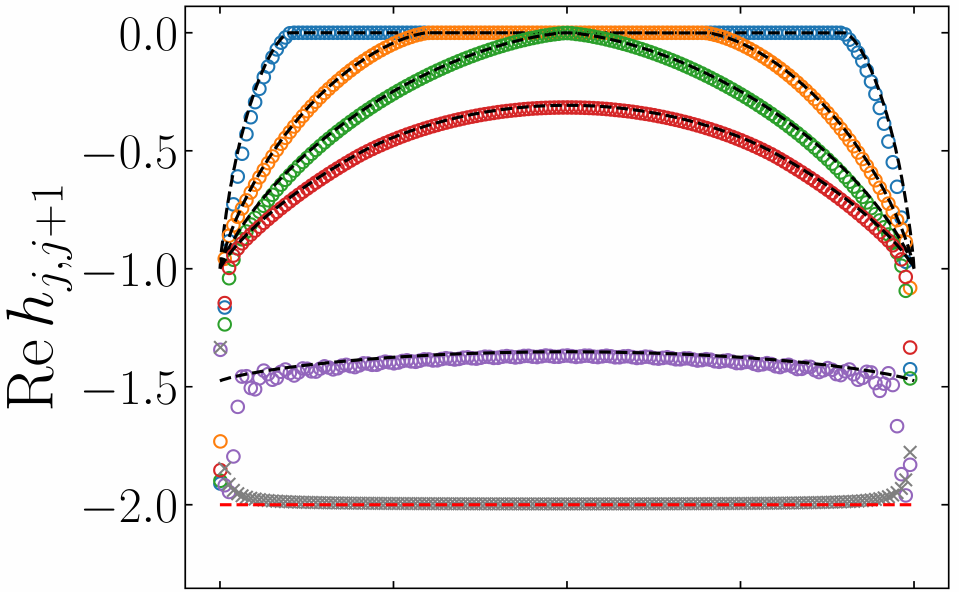}
\includegraphics[width=0.38\textwidth]{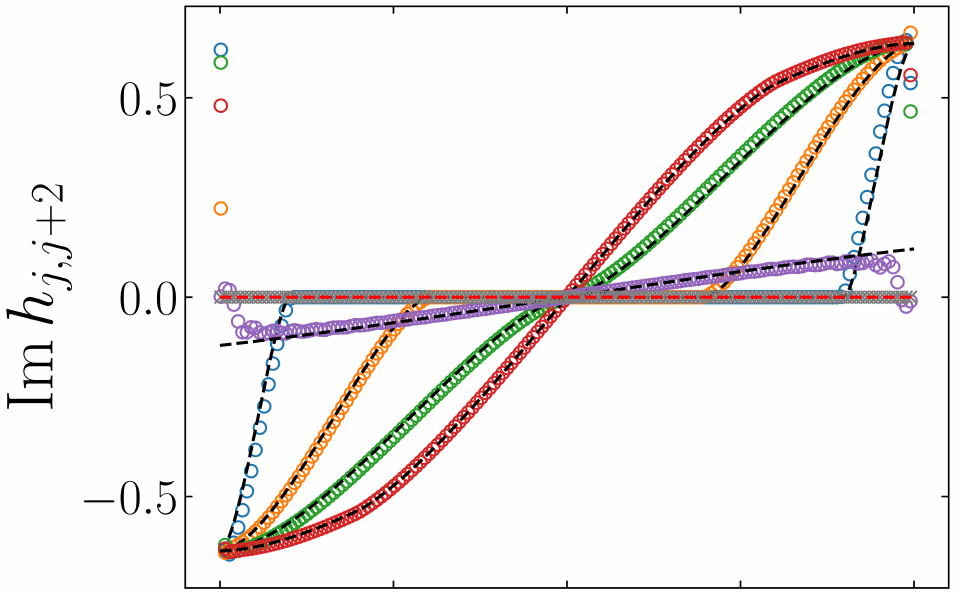}\\
\includegraphics[width=0.38\textwidth]{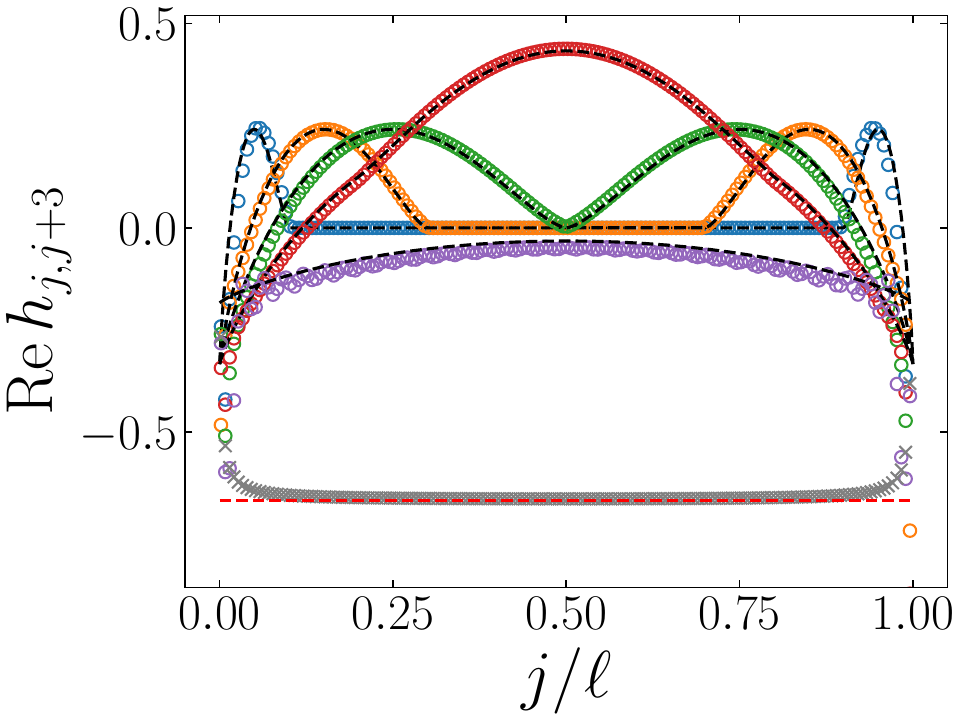}
\includegraphics[width=0.38\textwidth]{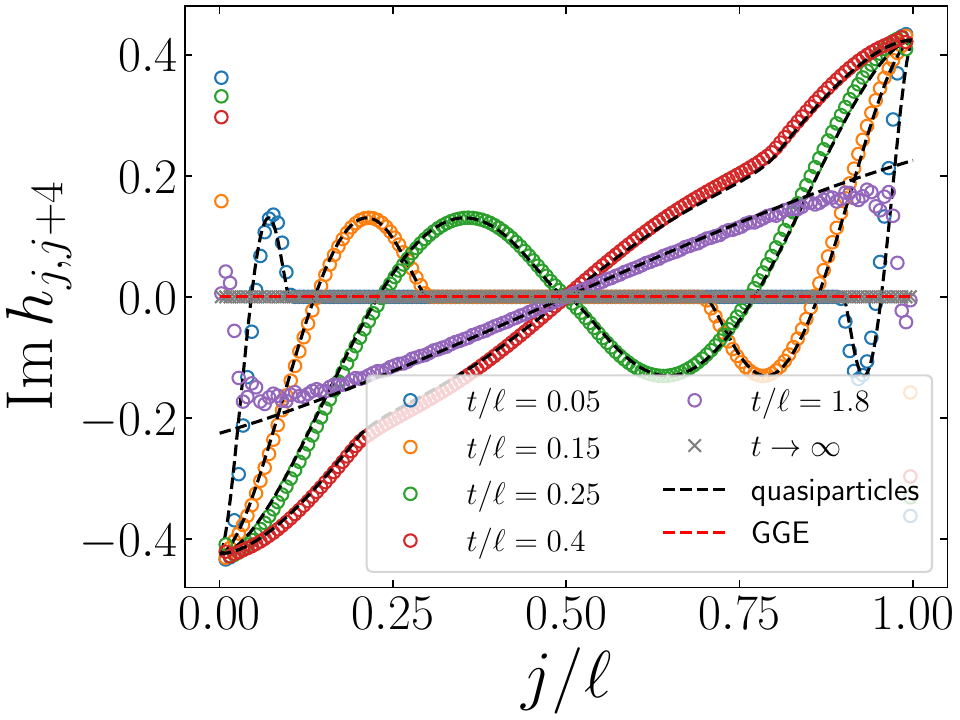}
    \caption{EH couplings $h_{j, j+z}$ between fermions at distances $z = 1, 2, 3, 4$ after a quench from the dimer state.
    We study an interval of length $\ell = 800$ at different times $t$ after the quench.
    The symbols are obtained from \cref{eq:peschel} using the correlation matrix~\eqref{eq:CAdimer}, while the dashed black lines are the kernels $\mathcal{K}_R(j/\ell,z)+\mathcal{K}_L(j/\ell,z)$ of the QPP prediction~\eqref{eq:EHspace} for different distances $z$.
    The couplings are real for odd distances $z$, while they are imaginary for even $z$.
    For $t \to \infty$,  (gray crosses), the data match the GGE, cf.  \cref{eq:GGE} (red dashed line).
    We observe a perfect agreement, with minor deviations near the endpoints where the QPP is not expected to work.
    For times $t/\ell \leqslant 0.25$ (blue and orange symbols) the couplings are non zero only inside of two light-cones centered at the endpoints.
    For $t > 0.25\ell$, the two light-cones merge and the EH relaxes toward the GGE.
    }
    \label{fig:z}
\end{figure*}

We prepare our system in some pure initial state $\rho$ and then allow it to undergo unitary time evolution according to the Hamiltonian of the form $H=\sum_k\varepsilon_k c^\dag_kc_k$ where $c^\dag_k,c_k$ are canonical fermions with energy $\varepsilon_k$ such that $[H,\rho]\neq 0$.
According to the QPP, each point in space acts as a source of quasiparticles that propagate through the system at velocity $v_k=\partial_k\varepsilon_k$ spreading correlations.
Particle pairs entirely contained within $A$ (or $\bar A$) do not contribute to the entanglement between $A$ and $\bar A$, while shared pairs do. 
Hence, the post-quench RDM can be written as the tensor product of an entangling and of a pure part
\begin{equation}\label{eq:RDMt}
    \rho_A^{(t)} \approx \frac{e^{- K_{A,\text{ QP}}^{(t)}}}{Z_A} \otimes \rho_{\text{pure}}^{(t)}\,.
\end{equation}
Here, the tensor product is inherited from the quasiparticle structure of the Hilbert space in the following way.  For every quasiparticle residing within $A$ we check if its correlated partner is also inside $A$ or not.  If it is, then this pair does not generate entanglement and contributes to $\rho_{\rm pure}^{(t)}$.   If it is not, then this generates entanglement and the quasiparticle  contributes to  $K_{A,\text{ QP}}^{(t)}$.   
The main result of this paper is that in the scaling limit of large times $t$ and large subsystem sizes $\ell$, with fixed ratio $\ell/t$, the EH $K_{A,\text{ QP}}^{(t)}$ takes the quadratic form
\begin{equation}\label{eq:EHspace}
    K_{A,\text{ QP}}^{(t)} = \int_{0}^{\ell} \dd x \int \dd z \left[ \mathcal{K}_R(x, z) + \mathcal{K}_L(x, z) \right] c_{x}^\dagger c_{x-z}\,,
\end{equation}
where $\mathcal{K}_R$ and $\mathcal{K}_L$ are the kernels
\begin{align}
    \mathcal{K}_R(x,z) &= \int_{k>0} \frac{\dd k}{2\pi}\, \eta(k)\,\Theta\!\left ( \min(2v_k t, \ell) - x \right ) e^{\ii k z}, \label{eq:kernelR}\\
    \mathcal{K}_L(x,z) &= \int_{k<0} \frac{\dd k}{2\pi}\, \eta(k)\,\Theta\!\left ( \max(\ell + 2v_k t, 0) - x \right ) e^{\ii k z}, \label{eq:kernelL}
\end{align}
 $\Theta(x)$ is the Heaviside function,  $c^\dag_x,c_x$ are the real space fermion operators  and $\ell=|A|$.
Furthermore, we introduced
\begin{equation}\label{eq:EHspectr}
    \eta(k) = \log\!\left [ \frac{1-n(k)}{n(k)} \right ],
\end{equation}
where $n(k)=\Tr[\rho\, c^\dag_k c_k]$.
As a conserved quantity, $\eta(k)$ can be computed in the initial state without solving the dynamics. Therefore, this form of the EH clearly enables its reconstruction without needing to solve the dynamics.
We stress that the main property of the form~\eqref{eq:EHspace} is to be entirely determined by quadratic terms with a kernel that depends on the distance $z$ in a light-cone fashion, see also \cref{fig:z} for an explicit example. 

The majority of the rest of this Letter is devoted to deriving this result and presenting several checks, numerical and analytical, of its validity.  Before this however, we make some brief comments on the structure of~\eqref{eq:EHspace}. 
Immediately after the quench, no quasiparticles are yet shared between $A$ and its complement. Accordingly, the full Hilbert space is identified with the one describing pairs, both of which are inside $A$, while the other one is empty, and the RDM is entirely captured by $\rho_{\text{pure}}^{(t)}$.
As time grows, the quasiparticles pass through the entangling points moving between $A$ and $\bar A$.
As a result, the dimension of the Hilbert space of the shared pairs increases while that of the unshared quasiparticles decreases. At long times, the Hilbert space of the shared quasiparticles becomes that of the full subsystem while that of the unshared quasiparticles becomes the empty set.
The entangling part is governed by the kernels $\mathcal{K}_{L,R}(x,z)$ which can be interpreted as the contributions from the left $(L)$ and right $(R)$ moving quasiparticles which spread from the left and right edges of $A$ respectively. 
From the structure of \cref{eq:kernelR,eq:kernelL}, we see that for $t < \ell/(4 v_{\rm max})$, where $v_{\rm max}=\max(v_k)$, the growth happens inside light-cones centered in $0$ and $\ell$, as expected from the QPP. The real space structure of the EH is determined by the initial state through $\eta(k)$.

In \cref{fig:z} we plot the values of the nearest neighbor and beyond nearest neighbor terms of the entanglement Hamiltonian, $\mathcal{K}_{L}(x,z)+\mathcal{K}_{R}(x,z),\, z=1,\dots,4$ for a particular case (see below for details).
What is evident, is that each of the terms exhibits a light cone like spreading emitting from the subsystem edges.
Moreover we see that the quasiparticle structure has endowed the hopping terms with a parity effect such that at finite times hopping over an odd number of sites is even with respect to reflection about the subsystem center and real, while hopping over an even number of sites is odd with respect to this and imaginary.
In the long time limit only odd site hopping terms remain.

\paragraph*{Derivation.}

\begin{figure}[t]
\includegraphics[width=.45\textwidth]{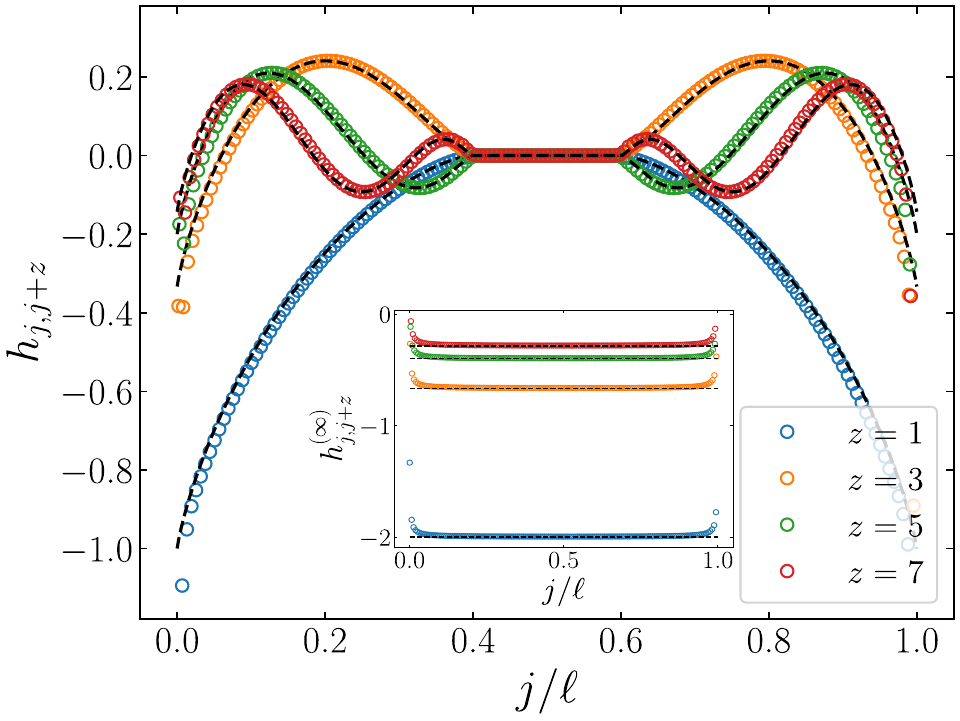}
    \caption{EH couplings $h_{j, j+z}$ after a quench from the dimer state.
    We consider an interval of length $\ell = 800$ at time $t/\ell = 0.2$.
    In the main plot we report the couplings (obtained using \cref{eq:peschel,eq:CAdimer}) for different (odd) distances $z$.
    The black dashed lines are the kernels $\mathcal{K}_R(j/\ell,z)+\mathcal{K}_L(j/\ell,z)$ of the prediction~\eqref{eq:EHspace}.
    The couplings present an oscillatory behavior with a peak which is slowly decreasing as the distance $z$ increases.
    The inset reports the couplings $h_{j, j+z}^{(\infty)}$ in the GGE which are approached at large times, for the same value of $z$ in the main plot.
    We see that the EH at intermediate time has a significantly different structure from the asymptotic value, demonstrating that the large time limit is reached in a highly non-trivial way.
    }
    \label{fig:diffZ_t02}
\end{figure}

As a starting point of our computation, we assume that at $t=0$ the initial state is a squeezed state of the form~\cite{ppv-17}
\begin{equation}\label{eq:squeeze}\begin{split}
    \ket{\psi^{(0)}} &= \mathcal{N} \exp\!\left \{ \ii \int_{k>0} \frac{\dd k}{2\pi}\, \mathcal{M}(k)\, c^\dagger_k c^\dagger_{-k} \right \} \ket{0}\,,
\end{split}\end{equation}
where $\ket{0}$ is the vacuum state of the post-quench Hamiltonian, $\mathcal{N}$ is the normalization and $\mathcal{M}(k)$ is some odd function whose particular form is unimportant.
In states of the form~\eqref{eq:squeeze}, the population $n(k)$ of the fermionic modes is
\begin{equation}\label{eq:occupation}
    n(k) = \frac{\mathcal{M}(k)^2}{1+\mathcal{M}(k)^2}\,.
\end{equation}
The choice of a state of the form~\eqref{eq:squeeze} greatly simplifies the derivation, since its pair structure, in which only modes of opposite momenta are entangled, can be read directly from \cref{eq:squeeze}.
Nevertheless, the final result is more general than the stated derivation, describing all free-fermionic quenches in which the initial state leads to the emission of entangled pairs of quasiparticles.
Moreover it can  be straightforwardly extended to account for states which have a more intricate quasiparticle structure~\cite{trc-24}.
As an example, in the next section we will study numerically a quench from the dimer state, whose scaling behavior, as can be seen from \cref{fig:z,fig:diffZ_t02}, is faithfully captured by our prediction, despite the initial state not being of the form~\eqref{eq:squeeze}.

To show that the RDM follows a QPP, in the spirit of Ref.~\cite{bfpc} we employ a hydrodynamic description.
We introduce hydrodynamic cells of length $\Delta$ much larger than the lattice spacing $a$ (or any other short distance cutoff) but smaller than the typical length $\ell$ of our system $a \ll \Delta \ll \ell$, and we impose periodic boundary conditions at the edges of the cell.
Quasiparticles are defined as wave-packets localized inside a hydrodynamic cell, obtained by performing a Fourier transform only inside the cell~\cite{bfpc}
\begin{equation}
\label{eq:quasiFourier}
    b^\dagger_{x, k} = \frac{1}{\Delta} \int_0^\Delta \dd y\, e^{-\ii k y}\, c^\dagger_{x+y}\,,
    \qquad 
    c^\dagger_{x+y} = \sum_{k} e^{\ii k y}\, b^\dagger_{x, k}\,,
\end{equation}
where $x$ labels the fluid cell and $y$ the position inside the cell. 
We now make the fundamental assumption that the correlations in the initial state~\eqref{eq:squeeze} decay fast enough with distance in real space.
If this holds, by taking hydrodynamic cells larger than the correlation length $\Delta \gtrsim \xi$, we can neglect correlations between different cells and we can approximate the initial state as a tensor product over the cells
\begin{equation}
    \rho^{(0)} = \ket{\psi^{(0)}} \!\!\bra{\psi^{(0)}} \approx \bigotimes_{k>0} \bigotimes_{x_0} \rho_{k,x_0}^{(0)}\,,
\end{equation}
where $x_0$ labels the hydrodynamic cell and $\rho_{k,x_0}^{(0)}$ is the density matrix of a single pair of quasiparticles~\cite{bfpc}
\begin{multline}\label{eq:rho0QP}
  \rho_{x_0,k}^{(0)} = n(k) \hat{n}_{x_0,k}\hat{n}_{x_0,-k}
    \\+(1-n(k))  ( 1-\hat{n}_{x_0,k})(1- \hat{n}_{x_0,-k}  ) \\
    + \ii \sqrt{n(k)(1-n(k))}  (b^\dagger_{x_0,k} b^\dagger_{x_0,-k} - b_{x_0,-k} b_{x_0,k}  )\,.
\end{multline}
where $\hat{n}_{x_0,k}=b^\dag_{x_0,k}b_{x_0,k}$.
In this approximation, only pairs of quasiparticles with opposite momenta and occupying the same cell are entangled with each others~\cite{bfpc}.

At time $t>0$, we evolve the state with the quenching Hamiltonian discussed above.
If the hydrodynamic cells are large enough compared with the lattice spacing, the diffraction of the wave-packet is negligible and under unitary evolution the quasiparticles move ballistically with group velocity $v_k$
\begin{equation}\label{eq:evolution}
    e^{\ii H t} b_{x_0,\pm k} e^{-\ii H t} = b_{x_0\pm v_k t, \pm k}\,.
\end{equation}
The density matrix~\eqref{eq:rho0QP} of a single pair of quasiparticles which originated in $x_0$, at time $t$ becomes~\cite{bfpc}
\begin{multline}\label{eq:rhoTQP}
    \rho^{(t)}_{x_0,k} = n(k)\, \hat{n}_{x_0+v_kt,k}\hat{n}_{x_0-v_kt,-k}\\
    + (1-n(k)) (1-\hat{n}_{x_0+v_kt,k}) (1-\hat{n}_{x_0-v_kt,-k})\\
    + \ii \sqrt{n(k)(1-n(k))} (b^\dagger_{x_0+v_k t,k} b^\dagger_{x_0-v_k t,-k} \\
    - b_{x_0-v_k t,-k} b_{x_0+v_k t,k}  )\,.
\end{multline}
We now compute the RDM of an interval $A = [0, \ell]$.
Thanks to the structure of the density matrix as the product states of quasiparticle pairs, we only need to study the RDM of a single pair.
To fix the ideas, consider a pair starting at position $x_0\in A$. 
At the beginning of the evolution, both quasiparticles are contained in the interval.
In this case, tracing out the degrees of freedom in the complement $\bar{A}$ has no effect on the RDM of the pair, which remains pure and equal to \cref{eq:rhoTQP} and does not contribute to the entanglement entropy.

At a later time, the, e.g., left-moving quasiparticle escapes the interval $x_0-v_k t < 0$, while the right-moving one is still inside $0 < x_0 + v_k t < \ell$.
Tracing out over $\bar{A}$ we obtain the RDM of the right-moving fermion, which is mixed and equal to 
\begin{multline}\label{eq:RDMtpair}
    \rho_{A, x_0, k}^{(t)} = n(k)\,\hat{n}_{x_0+v_kt,k}
    + (1-n(k)) (1-\hat{n}_{x_0+v_kt,k})\,.
\end{multline}
By following the trajectories of the quasiparticles, we see that at time $t$ the quasiparticles that are shared between $A$ and its complement are the right-movers that originated at $x_0 \in [-v_k t, \min(v_k t, \ell-v_k t)]$
and the left-movers that started at $x_0 \in [\ell - \min(\abs{v_k} t, \ell-\abs{v_k} t), \ell+\abs{v_k} t]$.
Expressing everything in terms of the current position of the quasiparticle $x = x_0 + v_k t$ for right-movers and $x = x_0 - \abs{v_k} t$ for left movers, we have
\begin{equation}\label{eq:RDMt2}
    \rho_A^{(t)} \approx \rho_{A, R\text{ QP}}^{(t)} \otimes \rho_\text{pure}^{(t)} \otimes \rho_{A, L\text{ QP}}^{(t)}\,,
\end{equation}
where 
\begin{align}
    &\rho_{A,R\text{ QP}}^{(t)} = \bigotimes_{k>0} \bigotimes_{x = 0}^{\min(2v_k t, \ell)} \rho_{A, x, k}^{(t)}\,,\\
    &\rho_{A,L\text{ QP}}^{(t)} = \bigotimes_{k<0} \bigotimes_{\substack{x = \ell-\\\min(2|v_k| t, \ell)}}^{\ell} \rho_{A, x, k}^{(t)}\,,
\end{align}
are the mixed parts of the RDM and $\rho_\text{pure}^{(t)}$ is the pure part due to the non-shared pairs.

To find the entanglement Hamiltonian of the mixed part of \cref{eq:RDMt}, we rewrite \cref{eq:RDMtpair} as an exponential, using the property $\hat{n}_{x_0,k}^2 = \hat{n}_{x_0,k}$.
We obtain $\rho_A^{(t)} = e^{-K_{A,\text{ QP}}^{(t)}}/Z_A$, where $Z_A$ is a normalization and 
\begin{multline}\label{eq:EHQP}
    K_{A,\text{ QP}}^{(t)} = \int_{k>0} \frac{\dd k}{2 \pi} \int_{0}^{\min(2v_k t, \ell)} \hspace{-1cm}\dd x\, \eta(k)\, b_{x,k}^\dagger b_{x,k} \\
   +\int_{k<0} \frac{\dd k}{2 \pi}  \int_{\max(\ell -2 \abs{v_k} t, 0)}^{\ell} \hspace{-1.5cm}\dd x\, \eta(k)\, b_{x,k}^\dagger b_{x,k}\,,
\end{multline}
The final step is to express the EH~\eqref{eq:EHQP} in terms of fermions in real space $c_{x}^\dagger$, $c_{x}$.
This simply amounts to a change of basis, performed using the inverse Fourier transform in the cell~\eqref{eq:quasiFourier}.
Performing this, we arrive to the main result of this work, \cref{eq:EHspace}. An analogous quasiparticle expression can also be determined for $\rho^{(t)}_{\text{pure}}$ thereby completely fixing the full RDM.

\paragraph*{Analytic checks.}
At sufficiently large times after the quench, the system relaxes locally to a stationary state which, generically for a free model is a generalized Gibbs ensemble (GGE) that incorporates all conserved charges.
The Hilbert space of the pairs which are  completely inside $A$ is empty and in the RDM~\eqref{eq:RDMt}
the pure part no longer contributes, while the EH~\eqref{eq:EHspace} fully characterizes the RDM.
In this limit, the time dependence of $\mathcal{K}_{L,R}$ drops out and asymptotic value of the EH~\eqref{eq:EHspace} is 
\begin{equation}\label{eq:GGE}\begin{aligned}
    K_{A,\text{ QP}}^{(\infty)} 
    &= \int_{0}^{\ell} \dd x \int_{0}^{\ell} \dd y \left [ \int \frac{\dd k}{2\pi}\, \eta(k)\, e^{\ii k(x-y)} \right ] c_{x}^\dagger c_{y}\\
    &= \int \frac{\dd k}{2\pi}\, \eta(k)\, c_{k}^\dagger c_{k},
\end{aligned}\end{equation}
which yields the expected GGE and allows $\eta(k)$ to be interpreted as the Lagrange multiplier of the conserved charges $c^\dagger_k c_k$~\cite{ef-16}.
This shows that \cref{eq:EHspace} correctly describes the generalized thermalization of the subsystem after the quantum quench. 

The light-cone behavior of the EH and eventual relaxation is especially evident when the quenching Hamiltonian has a linear dispersion, $\varepsilon_k=v k$. In this case the quasiparticle velocity $v$ is independent of $k$ and the Heaviside functions can be taken outside of the mode integrals in~\eqref{eq:kernelR}, \eqref{eq:kernelL}. 
Moreover, if $\eta(k)=\beta \varepsilon_k$, as happens after a quench in CFT~\cite{cc-06}, we find that
 \begin{multline}
      K_{A,\text{ QP}}^{(t)} = \beta \int_{0}^{\min(2v t,\ell)} \dd x \int \dd z \int_{k>0} \frac{\dd k}{2 \pi}\, \varepsilon_k\, e^{\ii k z}\, c_{x}^\dagger c_{x-z} \\
      + \beta \int_{\max(\ell -2v t, 0)}^{\ell} \dd x \int \dd z \int_{k<0} \frac{\dd k}{2 \pi}\, \varepsilon_k\, e^{\ii k z}\, c_{x}^\dagger c_{x-z} \\
      = \beta \left [ \int_{0}^{\min(2v t,\ell)} \dd x\, T(x) + \int_{\max(\ell -2v t, 0)}^{\ell} \dd x\, \overline{T}(x)\right ],
      \label{cfth}
\end{multline}
where $T$ and $\overline{T}$ are the right and left moving stress-energy operators.
\cref{cfth} agrees with previous results for the post-quench entanglement Hamiltonian in a CFT~\cite{ct-16,wrl-18}. Note that in this case the system relaxes locally to a Gibbs state in which only the Hamiltonian appears~\cite{c-16} and accordingly the EH has only short ranged hopping terms. This however, is not generically the case and $K_{A,\text{ QP}}^{(t)}$ can have a more intricate structure depending on both the initial state as encoded in $\eta(k)$ and $v_k$.

As written in~\eqref{eq:EHQP} it is straightforward to recover the QPP prediction for the growth of the R\'enyi entanglement entropy $S^{(\alpha)}_A=\frac{1}{1-\alpha}\log\Tr_A \left(\rho_A\right)^\alpha$. To achieve this we note that the tensor product structure of the RDM means we can compute the contribution of each quasiparticle, $b_{x,k}^\dag$ and that the entangling part for each of these takes the form of a generalized Gibbs state. Equating the R\'enyi entanglement entropy with the Fermi-Dirac R\'enyi entropy of this state and summing over all  quasiparticle contributions we find
\begin{equation}
    S^{(\alpha)}_A(t)=\int \frac{\dd k}{2\pi}\,\min(2|v_k| t,\ell)\,h_\alpha (n(k))\,,
\end{equation}
where $h_\alpha(x)=\frac{1}{1-\alpha}\log(x^\alpha+(1-x)^\alpha)$. Along similar lines and by including also the form of the pure part one can reproduce all previous QPP predictions for correlation functions~\cite{cef-12}, full counting statistics~\cite{bcckr-22} and symmetry resolved entanglement measures~\cite{bcckr-22,pbc-20,pbc-21b}.

\paragraph*{Numerical analysis.}

To verify the validity of our entanglement Hamiltonian in \cref{eq:EHspace}, we study a quench from the dimer state $\ket{D}=\prod_{j=1}^{L/2}(c_{2j}^{\dagger}+c_{2j-1}^{\dagger})\ket{0}/2^{L/4}$
to the hopping Hamiltonian
 $   H = - \frac{1}{2} \sum_{i} c_{i}^\dagger c_{i+1} + \text{h.c.}$. 
This quench has been studied in Ref.~\cite{ep-07}, where it was found that the two-point correlation matrix at time $t$ is given by~\cite{ep-07,gat-19,pbc-21b}
\begin{equation}\label{eq:CAdimer}
    C_{i,j}^{(t)} = C_{i,j}^{(\infty)} + \ii \frac{i-j}{4 t} e^{-\ii \frac{\pi}{2} (i+j)} J_{i-j}(2t),
\end{equation}
where $J_{\nu}(z)$ is Bessel's function and we have introduced the asymptotic value of the correlation matrix
{$ C_{i,j}^{(\infty)}=\delta_{i,j}/4+(\delta_{i,j-1} + \delta_{i,j+1})/2$}.
From the correlation matrix $C_A$ restricted to the subsystem $A$ it is possible to directly compute the entanglement Hamiltonian~\cite{Peschel2003,Peschel2004,Peschel2009,Peschel2012}.
Writing the lattice entanglement Hamiltonian as
\begin{equation}
    K_A = \sum h_{i,j} c_{i}^\dagger c_{j}\,,
\end{equation}
the single particle entanglement Hamiltonian $h$ is given by~\cite{Peschel2009}
\begin{equation}\label{eq:peschel}
    C_A = \frac{1}{1+e^h}\,.
\end{equation}
In order to compare the result of \cref{eq:peschel} with our prediction in \cref{eq:EHspace}, however, we need to take into account that at finite time \cref{eq:EHspace} only describes the low lying part of the entanglement spectrum.
Since according to \cref{eq:peschel} the higher part of the entanglement spectrum is given by the eigenvalues of $C_A$ that are close to either $0$ or $1$, we introduce a cut-off on the spectrum, projecting out all the eigenvalues of \cref{eq:CAdimer} which are smaller than $10^{-4}$ or larger than $1-10^{-4}$. We achieve this by first diagonalizing $C_A$ then projecting on to the desired eigenspace after which we perform the inverse transformation and extract $h_{j,j+z}$ using~\eqref{eq:peschel}.

Applying \cref{eq:peschel} to the asymptotic value of the correlation matrix, we can also immediately compute the large time EH, which agrees with the expected GGE.
In the $t \to \infty$ limit, the full entanglement spectrum is described by the GGE, therefore in this case we do not impose any cut-off on the spectrum.
The time-dependent numerical results for the EH are reported in \cref{fig:z,fig:diffZ_t02} showing a perfect match with the QPP prediction.

\paragraph*{Conclusions and Outlook.}
In this Letter, we derived a quasiparticle prediction, \cref{eq:EHspace}, for the EH after a global quantum quench in any system of non-interacting fermions. Remarkably, this simple form effectively captures the complex structure of the EH.

Our result paves the way for many intriguing generalizations. One key advantage of the QPP is its adaptability for calculating the entanglement entropy of disjoint intervals~\cite{cc-05,ac-19}, as well as more complex quantities like the negativity~\cite{ctc-14,ac-18b}.
Generalizing these results to an operator level, such as calculating the entanglement Hamiltonian of disjoint intervals or the negativity Hamiltonian~\cite{mvdc-22,rmtc-22,rmc-23}, would be extremely interesting.
Additionally, the QPP applies also for certain dissipative systems~\cite{ac-21,ca-22,ac-22,ca-24}, where also we could think of adapting our derivation.
Furthermore, the QPP has been recently combined with dimensional reduction~\cite{Chung-2000,mrc-20} to describe the entanglement entropy~\cite{yac-24} and asymmetry~\cite{yac-24b} in higher dimensional non-interacting models. The same could be done also for the EH.
A significant challenge is generalizing our result to interacting integrable models, where the QPP breaks down in the calculation of R\'enyi entanglement entropy~\cite{bka-22} and charged moments~\cite{bcckr-22}.

From a speculative perspective, it is well known that the primary difficulty in numerically studying with tensor networks the time evolution after a quench lies in the rapid growth of entanglement entropy~\cite{tns}. 
However, once the EH is known, its properties (e.g. the spectrum) may be extracted by means of equilibrium simulations (like quantum Monte Carlo) even when there is a volumetric scaling of the entropy. 
Furthermore, as recently proposed~\cite{dvz-17,kbevz-21,var-2}, the EHs may be engineered in cold-atom and trapped-ion setups to experimentally access the spectrum, even in the absence of a viable numerical algorithm.

\paragraph*{Acknowledgments}
We thank V. Alba, F. Ares, L. Capizzi, O. Castro-Alvaredo, B. Doyon, G. Di Giulio, J. Dubail, S. Murciano, and P. Zoller for fruitful discussions. 
All authors acknowledge support from ERC under Consolidator Grant number 771536 (NEMO). 
FR acknowledges support from the project ``Artificially devised many-body quantum dynamics in low dimensions -- ManyQLowD'' funded by the MIUR Progetti di Ricerca di Rilevante Interesse Nazionale (PRIN) Bando 2022 -- Grant 2022R35ZBF.

\end{document}